\documentclass[a4paper,fleqn,usenatbib]{mnras}

\usepackage{mathptmx}

\usepackage[T1]{fontenc}
\usepackage{ae,aecompl}
\usepackage{amsmath}

\usepackage{graphicx}	
\usepackage{amsmath}	
\usepackage{amssymb}	
\usepackage{rotating}

\newcommand{\around}{$\sim$}

\newcommand{\orcid}[2]{\href{http://orcid.org/#2}{#1}}

\title[Ultraluminous Quasars At High Redshift Show Evolution In Their Radio-Loudness Fraction]{Ultraluminous Quasars At High Redshift Show Evolution In Their Radio-Loudness Fraction In Both Redshift And Ultraviolet Luminosity}

\author[Lah et al.]
{\orcid{Philip Lah}{0000-0001-6841-6553}$^{1}$\thanks{E-mail: philip.lah@anu.edu.au},
\orcid{Christopher A.~Onken}{0000-0003-0017-349X}$^{1}$, 
\orcid{Ray P.~Norris}{0000-0002-4597-1906}$^{2,3}$,
and \orcid{Francesco D'Eugenio}{0000-0003-2388-8172}$^{4,5}$\\
$^{1}$Research School of Astronomy and Astrophysics, Australian National University, Canberra, ACT 2611, Australia\\
$^{2}$Western Sydney University, Locked Bag 1797, Penrith, NSW 2751, Australia \\
$^{3}$CSIRO Space \& Astronomy, P.O.~Box 76, Epping, NSW 1710, Australia \\
$^{4}$Kavli Institute for Cosmology, University of Cambridge, Madingley Road, Cambridge, CB3 0HA, UK\\
$^{5}$Cavendish Laboratory, University of Cambridge, 19 JJ Thomson Avenue, Cambridge, CB3 0HE, UK\\
}

\date{Accepted XXX. Received YYY; in original form ZZZ}

\pubyear{2022}

\begin{document}
\label{firstpage}
\pagerange{\pageref{firstpage}--\pageref{lastpage}}
\maketitle


\begin{abstract}

{We take a sample of 94 ultraluminous, optical quasars from the search of over 14,486 deg$^2$ by \citet{onken22} in the range $4.4<$redshift$<5.2$ and match them against the Rapid ASKAP Continuum Survey (RACS) observed on the Australian Square Kilometre Array Pathfinder (ASKAP).  From this most complete sample of the bright end of the redshift $\sim 5$ quasar luminosity function, there are 10 radio continuum detections of which 8 are considered radio-loud quasars.  The radio-loud fraction for this sample is $8.5 \pm 2.9$ per cent.   \citet{jiang07} found that there is a decrease in {the} radio-loud fraction of quasars with increasing redshift and an increase with increasing absolute magnitude at rest frame 2500~\AA.  We show that the radio-loud fraction of our quasar sample is consistent with that predicted by \citet{jiang07}, extending their result to higher redshifts.}  

\end{abstract}





\section{Introduction}

Quasars were first discovered through their radio emission \citep{matthews63,schmidt63}.  However, it was soon found that the majority of quasars had little or no detectable radio emission \citep{sandage65}.  Quasars are often classified into two categories: radio-loud and radio-quiet, based on the ratio of their radio to optical flux density.  Some authors find a bimodal distribution for radio-loud and radio-quiet quasars \citep{kellermann89,miller90,visnovsky92,goldschmidt99,ivezic02,ivezic04,white07}.  Others find no evidence for such a distribution \citep{cirasuolo03,lacy01,singal11,balokovic12,singal13,macfarlane21}.

The spectral energy distributions of radio-loud and radio-quiet quasars are very similar \citep{elvis94,richards06,devries06,shang11,shankar16}, {differing} only in the X-ray  and  radio bands.    At low redshift the vast majority of radio-loud quasars are hosted by giant elliptical galaxies while radio-quiet quasars are found to be hosted by both elliptical and spiral galaxies \citep{floyd10,tadhunter16,rusinek20}.  At higher redshifts radio-loud quasars have higher star formation than radio-quiet quasars \citep{kalfountzou12} which is consistent with their role as the progenitors of the low-redshift high mass ellipticals.  Radio-loud quasars are found in denser environments as well as having much more massive dark matter halos and higher stellar masses than radio-quiet quasars  \citep{mandelbaum09,shen09,donoso10,wylezalek13,rees16,retanamontenegro17}.

Quasars are thought to be triggered by high accretion associated with galactic mergers \citep{hopkins06}.  Most of the radio-loud quasars, and some  radio-quiet quasars, are thought {to be} due to mergers of elliptical galaxies with disk galaxies while the remaining radio quiet quasars are due to the mergers of two disk galaxies \citep{shen09,bessiere12,treister12}.  The dependency on the type of galaxy merger with the radio properties of the quasar can be tested by comparing the fraction of radio-loud quasars with the cosmic history of the different types of mergers.  The fraction of elliptical-spiral mergers decreases with increasing redshift while the fraction of spiral-spiral mergers increases with redshift as seen in semi-analytical models \citep{khochfar03} and in observations \citep{lin08}.  This suggests that the radio-loud fraction would decrease with increasing redshift.

Indeed, many authors have shown that the radio-loud fraction does decreases with increasing redshift \citep{peacock86,miller90,schneider92,visnovsky92,lafranca94}, although  some authors have shown no evolution with redshift \citep{goldschmidt99,stern00,ivezic02}.  Many authors have shown that the radio-loud fraction increases with optical luminosity \citep{visnovsky92,padovani93,goldschmidt99}, while a few authors find no such trend,\citep{stern00,ivezic02}, and a few authors show  the radio-loud fraction increasing with luminosity\citep{hooper95,bischof97}. {These early results were often complicated by small sample sizes and the complex interplay of redshift, luminosity, and survey flux limits.}

{Of particular relevance to this paper is the work by \citet{jiang07} where they used 30,000 optically selected quasars from the Sloan Digital Sky Survey matched to the Faint Images of the Radio Sky at Twenty-Centimeters (FIRST) radio survey to examine the properties of the radio fraction of quasars.  They found that the radio-loudness fraction of quasars is a function of both redshift and rest frame 2500~\AA\ luminosity with the fraction decreasing with redshift and increasing with luminosity.  The work of \citet{jiang07} did not extend out to the redshifts probed by our work reaching only out to redshift \around 4.6. The conclusions of \citet{jiang07} were supported by the later work of \citet{kratzer15,rusinekabarca21}.}

\citet{yang16} found for luminous quasars at $4.7<$redshift$<5.4$ that the {radio-loud fraction} may evolve with optical luminosity but that the fraction may not decline as rapidly with increasing redshift as measured by \citet{jiang07}.  \citet{banados15} in a sample of quasars at redshift$> 5.5$ found that there appeared to be no evolution in the radio-loud fraction at these redshifts as does \citet{liu21} for a similar redshift sample.

Ultraluminous quasars at high redshift are extremely rare but these objects are of particular interest as  the early universe is the era in which super massive black holes undergo their most dramatic and least explained growth.  Ultraluminous quasars are difficult to find as the {candidate lists} are swamped by the tail end distribution of cool, red stars from within our own Galaxy.  To find these rare objects \citet{onken22} used photometry from the SkyMapper Southern Survey (SMSS) Data Release 3 \citep{wolf18,onken19}, the 2 Micron All-Sky Survey (2MASS) \citep{skrutskie06}, the VISTA Hemisphere Survey (VHS) Data Release 6 \citep{mcmahon13}, the VISTA Kilo-degree Infrared Galaxy (VIKING) Survey Data Release 5 \citep{edge13},  AllWISE \citep{cutri21} and the CatWISE2020 Catalog \citep{marocco21}.  {(VISTA is the Visible and Infrared Survey Telescope for Astronomy and \emph{WISE} is the \emph{Wide-field Infrared Survey Explorer}).}  In addition they used proper motions from the {\it Gaia} Data Release 3 \citep{gaiacollaboration21} to remove stars from their candidate lists.  Spectroscopic {follow-up} was then done with the {Australian National University (ANU)} 2.3m telescope to confirm the targets as quasars and to obtain precise redshifts.  This yielded a sample of redshift $\sim5$ quasars with unprecedented completeness at the bright end.  

In this paper we have matched the quasars from \citet{onken22} to the new Rapid ASKAP Continuum Survey (RACS) \citep{mcconnell20,hale21} to measure the evolution of the radio-loud fraction for these objects,  as seen in previous work  \citep[e.g][]{jiang07}.  

In Section~\ref{Analysis} the quasar data from the optical and infrared combined with the radio data is analysed {to determine the radio-loud fraction of the sample}.  In Section~\ref{Sample_Completeness} the completeness of the sample in both radio and optical is examined.  In Section~\ref{Results} we discuss the radio-loud fraction, and its evolution with redshift and optical rest frame luminosity.  In Section~\ref{Conclusion} the conclusions from this work are summarised.


\section{Analysis}
\label{Analysis}

\citet{onken22} searched  for ultraluminous quasars at redshift greater than 4.4 over 14,486 deg$^2$ of the {whole sky} from Dec $< +2$~degrees, Galactic latitude |b|$>$15~degrees and excluding some regions around the Magellanic Clouds and other nearby galaxies in the local group as well as regions around bright stars.  {Quasar candidates were limited to those with a  SkyMapper z-band zPSF $< 19.5$ AB magnitude.  Cross-matches were made with the large-area surveys {\it Gaia}, WISE and VISTA and candidates with a neighbour within 5 arcseconds were removed.  The following selection criteria were used to remove contamination by cool, red stars and lower redshift quasars: 
\begin{equation}
\begin{split}
& 0.8 < G - Rp < 1.8 \\
& 1.8 < Bp,c - Rp \\
& 0.9 < gPSF - rPSF \\
& 0.7 < J - K < 1.8 \\
& 1.5 < J - W1 < 3 \\
& 0.2 < W1 - W2 < 1.1 \\
& 2.3 < W1 - W3 < 4.7 \\
& 0 < (J - K) + (Bp,c - Rp ) - (zPSF - J ) - 1.8 \\
& 0 < (J - W1) - 1.4(zPSF - J ) + 0.1 \\
& 0 < 0.6 - 0.5(rPSF - iPSF ) - (G - rPSF ) \\
\end{split}
\end{equation}
where gPSF, rPSF and zPSF are passbands from SkyMapper; G, Bp,c and Rp are passbands from {\it Gaia}; J, H and K are passbands from VHS and VIKING; W1 and W2 are passbands from CatWISE and W3 is a passband from AllWISE.  Bp,c is corrected for magnitude-dependent biases.  The next criterion was that the {\it Gaia} proper motions and parallaxes for the objects were consistent with zero within the errors  (2$\sigma$).}  The quasar candidates were then followed up with spectroscopic observations with the ANU 2.3m telescope. Known quasars from Milliquas v7.1 \citep{flesch15} in the search area were added to the quasar sample.  

This quasar list was matched to the radio continuum sources from Rapid ASKAP Continuum Survey (RACS) from \citet{hale21}.  RACS was observed at a frequency of 887.5~MHz with a bandpass of 288~MHz.  The RACS sample from \citet{hale21} covers a Declination range from $-$80~degrees to +30~degrees excluding Galactic latitudes |b|$<$5~degrees, resulting in a good match to the region covered by \citet{onken22}.  RACS has a resolution of 25~arcsec and a median RMS \around 0.3~mJy~beam$^{-1}$.  There are \around 2.1 million radio continuum sources in RACS.  

The sample of optical quasars from \citet{onken22} chosen to match to RACS was limited to a range $4.4<$redshift$<5.2$ where the sample is most complete.  A cut off in the SkyMapper z~band of AB magnitude $<18.7$ was used for which the sample was 78 per cent complete \citep{onken22}.  {The enhanced completeness relative to the study of \citet{yang16} ($\sim55$ per cent) is primarily due to the recent availability of {\it Gaia} astrometry, which allowed further exploration of the ($W1-W2$) colour space without suffering from overwhelming stellar contamination. Below this magnitude cut the sample incompletness increases greatly.  Incompleteness is calculated from the variation of Skymapper z-band number counts with magnitude, and spectroscopic followup of candidate quasar objects.} This gave an optical quasar sample of 94 quasars.  {These optical quasars were matched to the objects in RACS with a maximum matching radius of 10~arcseconds}.  There were 10 matched objects, with the maximum matching radius being 2.83~arcseconds, and most being less than 1-arcsec.    

{The definition for radio loudness used in this paper is  that used by \citet{jiang07}, which is based on  earlier work by \citet{stocke92}.  This differs from other definitions as it is defined in the ultraviolet rather than the optical (e.g.~at 4400\AA\ used {by} \citet{kellermann89} from the original paper on radio loudness).  The advantage of using the ultraviolet is that the observed wavelength for high redshift quasars  is lower, making it easier to measure.}  

The radio loudness has the form:

\begin{equation}
\rm R = \frac{f_{5000\, MHz}}{f_{2500\, \mbox{\normalfont\AA}}}
\end{equation}

Where $\rm f_{5000\, MHz}$ is the flux density of the quasar at rest frame 5000~MHz and $\rm f_{2500\, \mbox{\normalfont\AA}}$ is the flux density at rest frame 2500~\AA.

At the redshifts of the quasar sample the redshifted 2500~\AA\ is close to the J and H bands (J, H and K are Vega magnitudes {and come from the VHS and VIKING surveys)}.  At the average redshift~=~4.67 for the quasar sample 2500~\AA\ is observed at 1.42~$\mu$m (J band is at 1.22~$\mu$m and H is at 1.63~$\mu$m).  To find the flux density at redshifted 2500~\AA\ the spectral index was calculated using J and H band {(after a minor correction for interstellar foreground extinction using the map from \citet{schlegel98})} using the  equation:

\begin{equation}
\rm \alpha = \frac{ \log(S1/S2)} { \log(\nu 1/\nu 2) }
\end{equation}

where S1 and S2 are the flux densities at the frequency of interest and $\nu 1$ and $\nu 2$ are the frequencies of interest.  The spectral index was then used to estimate the flux density at 2500~\AA\ including a 1+redshift cosmological correction.  In the cases where there was no J band {due to a lack of coverage by the surveys} (one object in the radio detections) a similar estimation was done using H band and K band (2.19~$\mu$m) instead and then extrapolating to the redshifted 2500~\AA\ magnitude.  In general where J, H and K band were all available, the value from his extrapolation from K band was close to the interpolated value from J and H band so doing the extrapolation is unlikely to significantly bias the results.  

At the redshifts of the quasar sample the redshifted 5000~MHz is close to the RACS observing frequency of 887.5~MHz.  At the average redshift=~4.67 for the quasar sample 5000~MHz is observed at 880~MHz.  To estimate the flux density at exactly 5000~MHz an $\alpha = -0.5$ was used for the quasars, which is the same as used by \citet{jiang07}, {so that our measurements would be as similar to theirs as possible. A spectral index of $\alpha = -0.5$ is typical for quasars  \citep[e.g.][]{ivezic04b}.}  


\begin{table*}
\centering
\caption{The properties of the quasars with radio continuum detections in RACS.  In the last two columns a `-' means that the object is outside the area covered by the surveys and and a `*' meant that the source was not detected in the survey. }
\label{tab:quasars}
\begin{tabular}{lcllllllllll} 
\hline

\ & SkyMapper    & \         & \         & \        & \   & \    & mag  & flux  & flux  & flux \\
\ & Southern Sky & RA        & DEC       & \        & z   & mag  & 2500 & RACS  & NVSS  & VLASS \\
R & Survey ID    & (degrees) & (degrees) & Redshift & mag & 2500 & abs  & (mJy) & (mJy) & (mJy) \\             

\hline

  5.2 & 035504.86-381142.5 &  58.77027 & -38.19514 &   4.545 &   17.44 &   19.04 &  -29.11 &    2.51 & 2.2 & 2.89 \\ 
  6.5 & 151443.82-325024.8 & 228.68260 & -32.84022 &   4.810 &   17.85 &   19.09 &  -29.20 &    3.24 & * & * \\ 
 15.9 & 232952.75-200038.7 & 352.46985 & -20.01085 &   5.030 &   18.45 &   19.92 &  -28.48 &    3.88 & * & * \\ 
 33.0 & 145147.04-151220.1 & 222.94603 & -15.20561 &   4.763 &   17.12 &   18.71 &  -29.56 &   22.94 & 28.5 & 44.82 \\ 
 55.9 & 013127.34-032059.9 &  22.86391 &  -3.34998 &   5.196 &   18.03 &   19.39 &  -29.10 &   23.19 & 31.4 & 49.78 \\ 
 93.9 & 013539.28-212628.2 &  23.91370 & -21.44118 &   4.940 &   17.74 &   19.21 &  -29.15 &   43.23 & 25.3 & 31.48 \\ 
207.4 & 043923.20-020701.6 &  69.84667 &  -2.11710 &   4.400 &   18.68 &   19.96 &  -28.10 &   41.37 & 43.3 & 50.58 \\ 
264.0 & 033951.43-473959.9 &  54.96432 & -47.66662 &   4.450 &   18.65 &   20.59 &  -27.51 &   29.97 & - & - \\ 
556.3 & 052506.17-334305.6 &  81.27573 & -33.71823 &   4.417 &   18.18 &   19.62 &  -28.45 &  152.02 & 188.3 & 104.79 \\ 
745.2 & 032444.28-291821.0 &  51.18452 & -29.30586 &   4.622 &   17.93 &   19.58 &  -28.61 &  224.03 & 236.5 & 161.21 \\ 

\end{tabular}
\end{table*}


The calculated radio loudness R along with the absolute AB magnitude at rest frame 2500\AA\  for the quasars with radio detections are shown in Table~\ref{tab:quasars}.  Here we define radio-loud as meaning R $> 10$, {which is the canonical value used by \citet{kellermann89} and others.}  Of the 10 quasars with radio detections 8 are radio-loud using this definition.  

Also in Table~\ref{tab:quasars} are the NRAO VLA Sky Survey (NVSS) \citep{condon98} and VLA Sky Survey (VLASS) \citep{gordon21} flux densities for the quasars where they are available.  {(NRAO is the National Radio Astronomy Observatory and VLA is the Very Large Array).} Both surveys only extend down to Declination $-$40~degrees so there is  incomplete overlap of the quasar sample.  The frequency of observation of NVSS is \around 1400~MHz and the survey has a resolution of 45~arcsecond.  NVSS has a flux density limit of \around 2.5~mJy~beam$^{-1}$.  {A maximum matching radius of 15~arcseconds was used to match NVSS to the optical quasars.}  The worst match with the quasar sample for NVSS is 13.5~arcseconds away, the next being 1.15~arcseconds.  The frequency of observation of VLASS is 3000~MHz and the survey has a resolution of 2.5~arcsecond.  VLASS has a median RMS sensitivity of 0.128~mJy~beam$^{-1}$.  {A maximum matching radius of 1.5~arcseconds was used to match VLASS to the optical quasars.}  The worst match with the quasar sample was a matching radius of  1.09 arcseconds.  For the RACS-detected sample, one object is not in the sky coverage of NVSS and VLASS and 2 objects were not detected by the surveys.  


\section{Sample Completeness}
\label{Sample_Completeness}


\begin{figure}
  \includegraphics[width=\columnwidth]{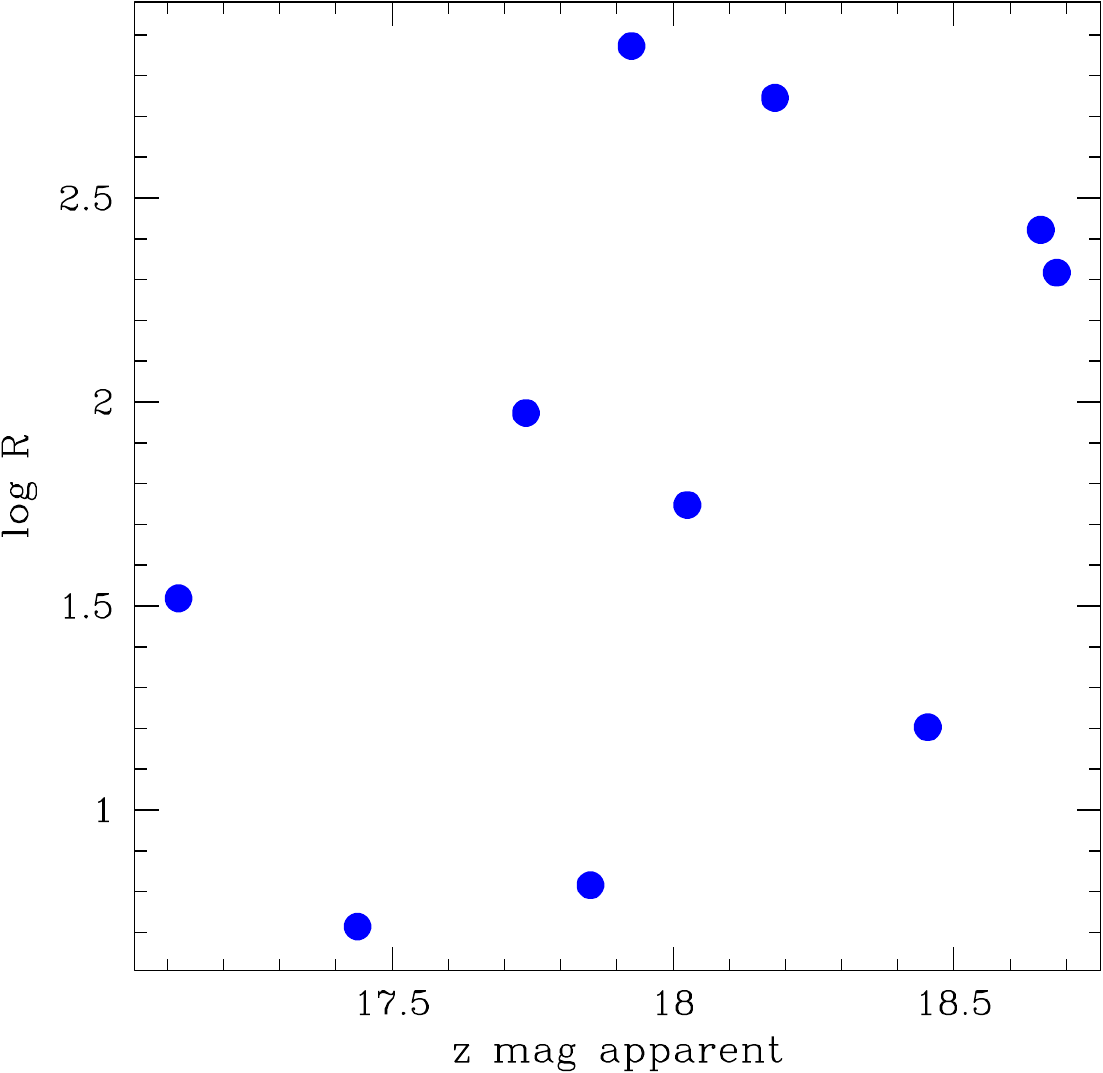}  
  \caption{The log of radio loudness of the radio detected quasars versus their apparent z band magnitude.  There is no correlation visible.}
  \label{fig:plot_radio_loud_mod}
\end{figure}


The optical quasar sample is 78 per cent complete to a SkyMapper z~magnitude $<18.7$ \citep{onken22}.  There is no correlation between radio loudness and z magnitude for the sample (see Figure~\ref{fig:plot_radio_loud_mod}).  Therefore in the missing 22 per cent of quasars it is reasonable to assume that there is the same ratio of radio-loud quasars as in the 78 per cent, {though it should be noted that we are dealing with small number statistics.}  So in order to reach 100 per cent completeness one can scale up both the radio detections and the total quasar sample by the same amount, which will leave the radio-loud fraction, the ratio of these two quantities, the same.


\begin{figure}
  \includegraphics[width=\columnwidth]{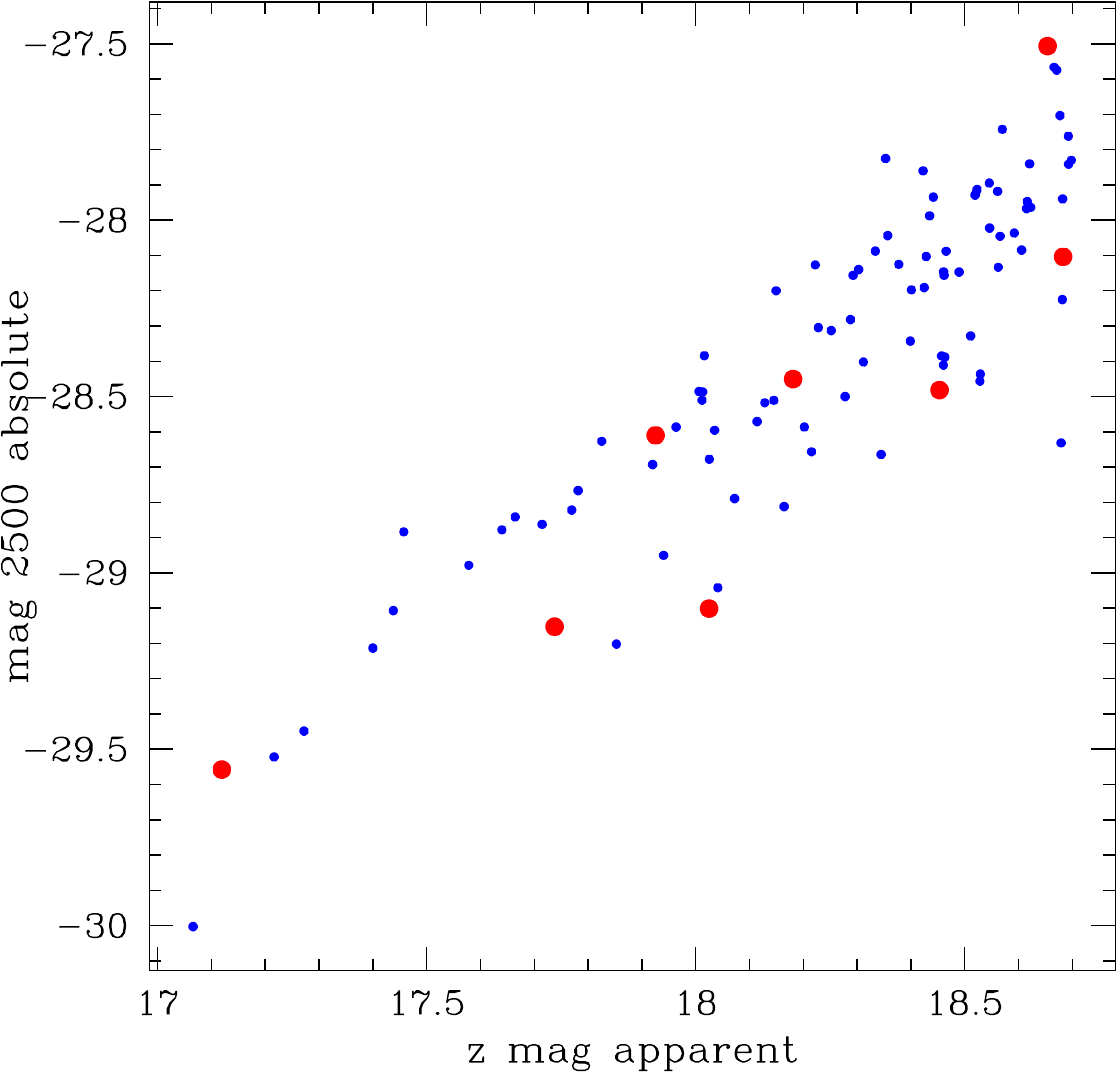}  
  \caption{The absolute magnitude at rest frame 2500\AA\ for all the optical quasars in the sample plotted against their z apparent magnitude.  The larger, red points are the radio-loud quasars.}
  \label{fig:plot_mag_redshift}
\end{figure}


Figure~\ref{fig:plot_mag_redshift} shows the absolute magnitude at rest frame 2500\AA\ for all the optical quasars in the sample plotted against their z apparent magnitude.  What one can immediately notice is that there is a strong correlation between absolute magnitude at rest frame 2500\AA\ and the apparent z magnitude.  Thus the cutoff at z~band magnitude of 18.7 can roughly be translated to a cut in the absolute magnitude at rest frame 2500\AA.  This means that we are not missing a large number of quasars by using this cut at z~band magnitude of 18.7 rather than a cut in absolute magnitude.  Only a few quasars may be missing which will have the effect of decreasing the measured radio-loud fraction by a small amount.  


\begin{figure}
  \includegraphics[width=\columnwidth]{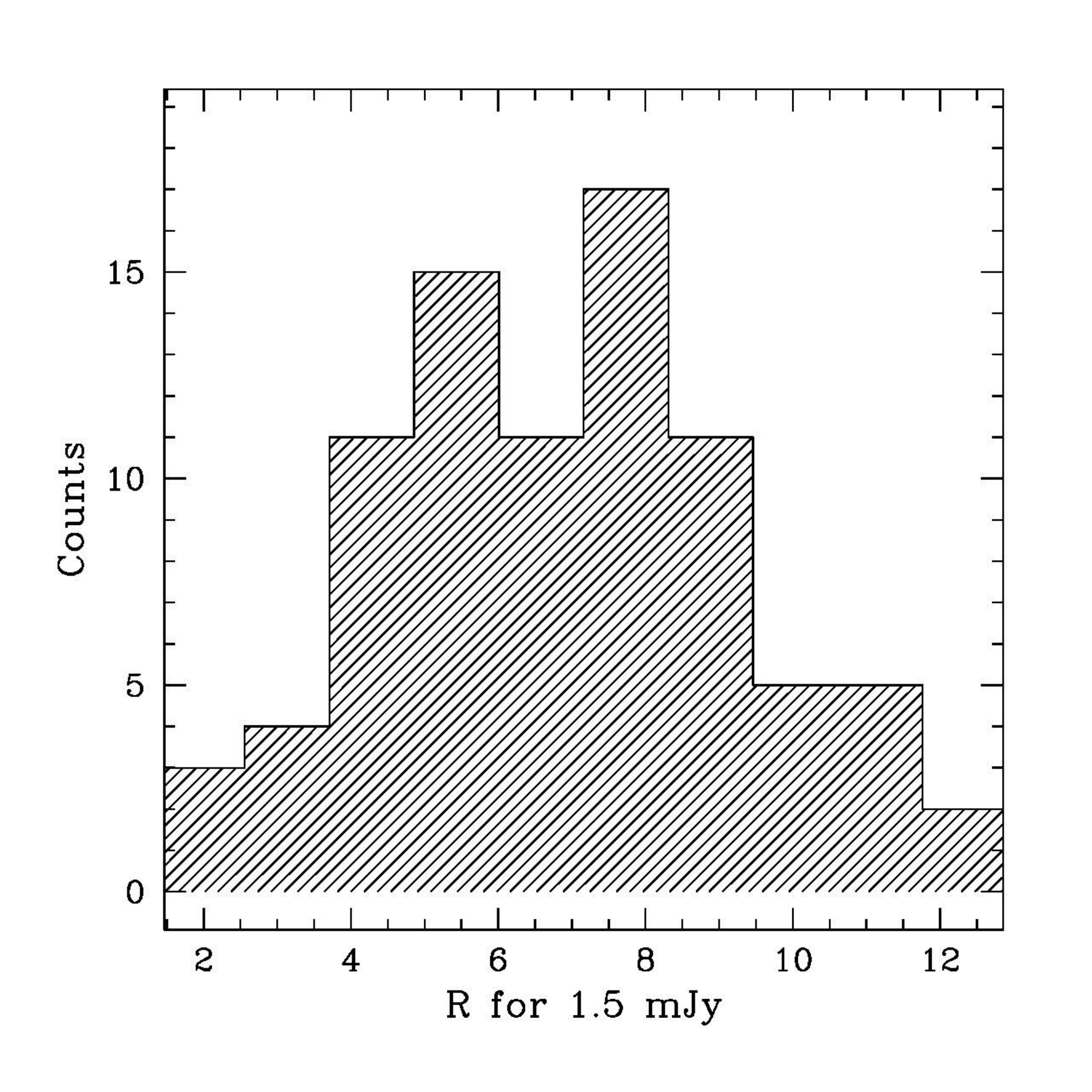}
  \caption{Radio loudness for the quasars {without radio detections} assuming that the quasars all have a RACS radio flux density of 1.5~mJy (5 times the median RMS of RACS).  Radio loud quasars have R$>$10.}
  \label{fig:hist_quasar_properties_R_2500}
\end{figure}

 
{Figure \ref{fig:hist_quasar_properties_R_2500} shows the radio loudness for the quasars without radio detections assuming that the quasars all have a RACS radio flux density of 1.5~mJy (5 times the median RMS of RACS).}  which would therefore  have been detected in RACS.  The fact that most of the quasars with this assumed flux density are below the cutoff for being radio-loud (R~=~10) means that we are not missing a large number of radio-loud quasars in our sample due to the RACS flux density limit. 


\section{Results}
\label{Results}


\begin{figure*}
  \includegraphics[width=\columnwidth]{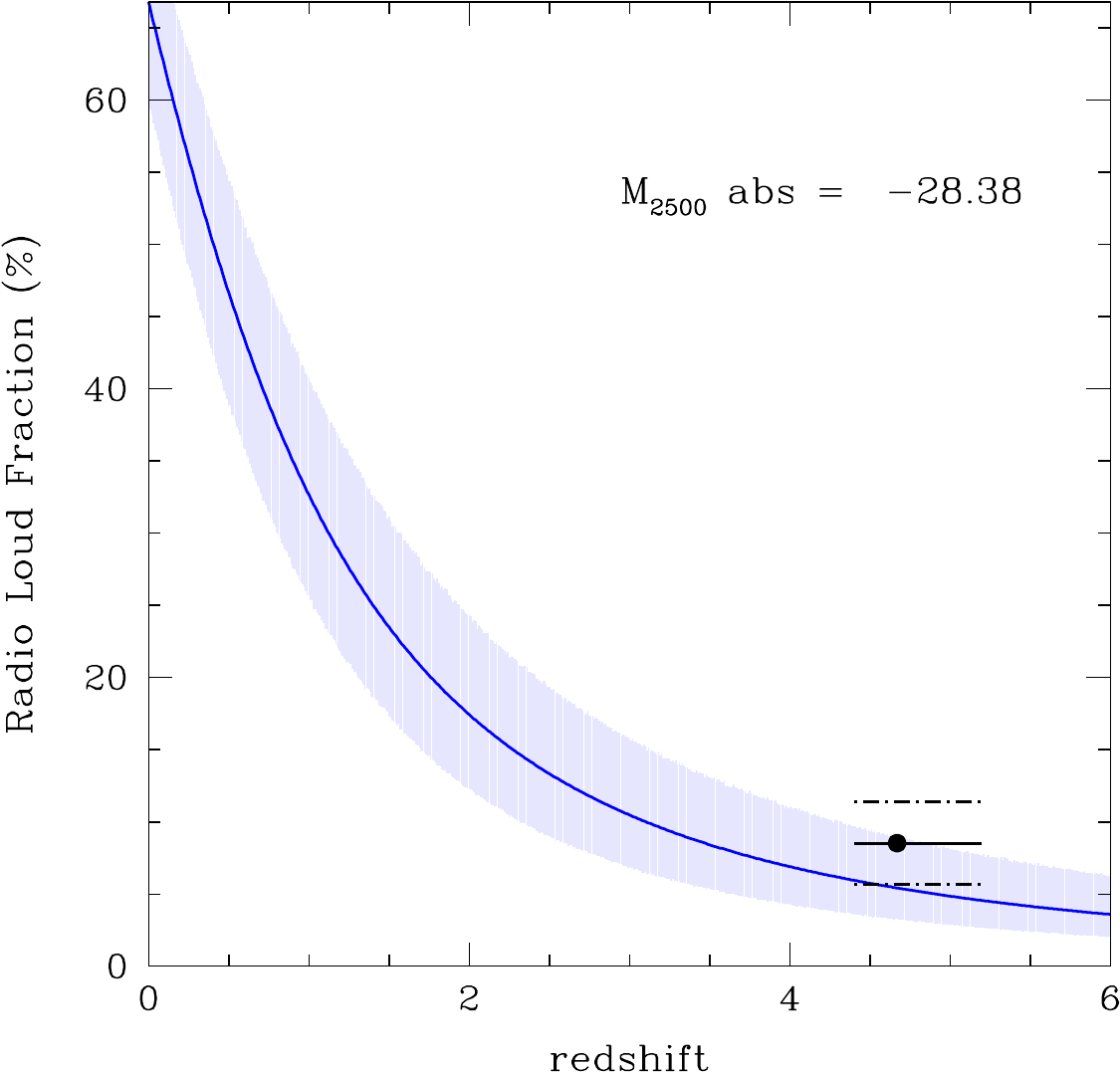}
  \includegraphics[width=\columnwidth]{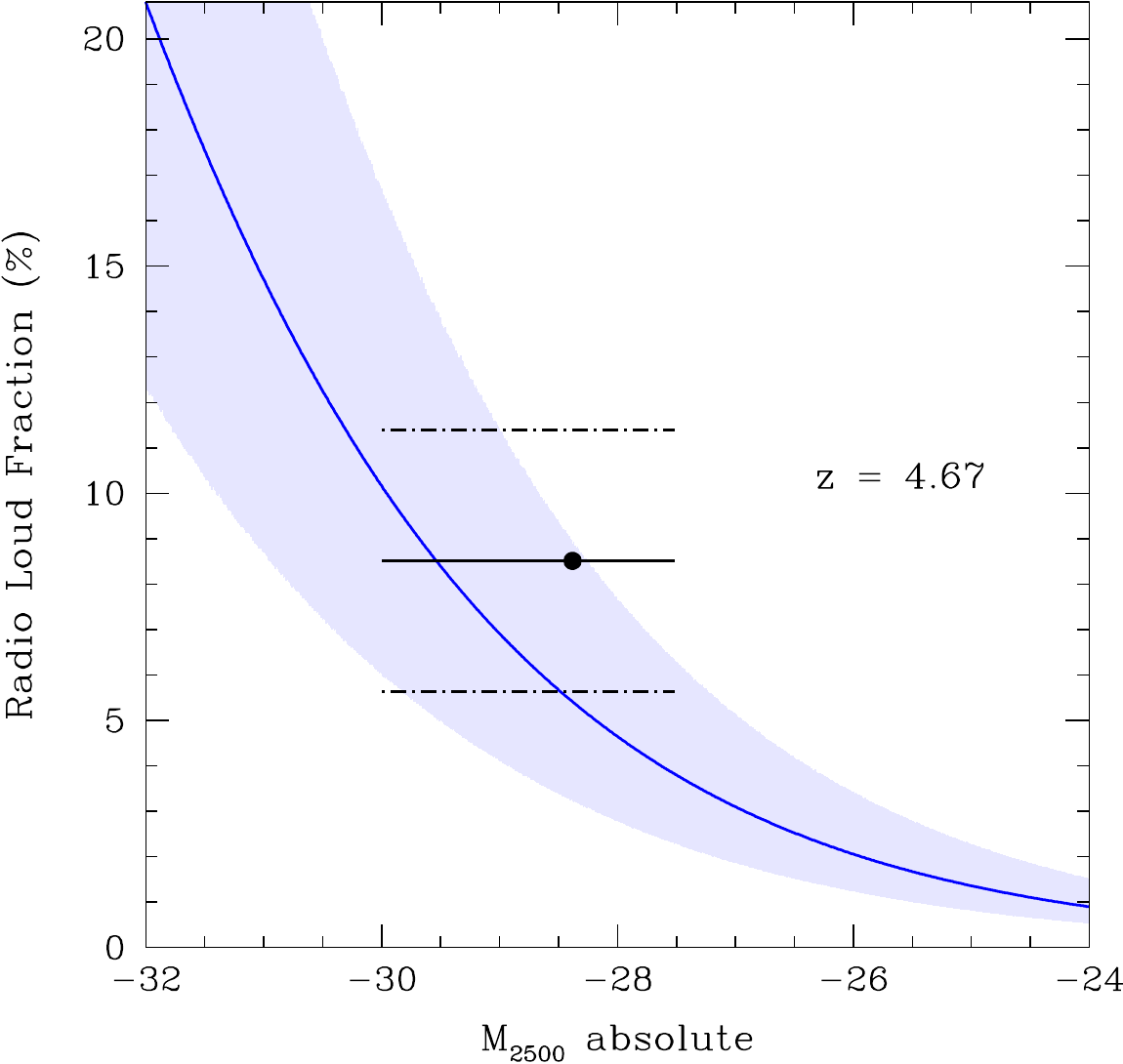}
  \caption{The left panel, using the equation of \citet{jiang07}, shows the radio-loud fraction as a function of redshift at $\rm M_{2500}=-28.38$ the average of the quasar sample.  The light blue is the error on this function.  The black point is the radio fraction from our quasar sample at the average redshift.  The black lines show the range of redshifts covered in our sample with the dotted lines showing the range of the error.
The right panel, using the equation of \citet{jiang07}, shows the radio-loud fraction as a function of absolute magnitude $\rm M_{2500}$ at  redshift=4.67 the average of the quasar sample.  The light blue is the error on this function.  The black point is the radio fraction from our quasar sample at the average $\rm M_{2500}$.  The black lines cover the range of $\rm M_{2500}$ covered in our sample with the dotted lines showing the range of the error.}
  \label{fig:plot_jiang07_mod}
\end{figure*}


The number of radio-loud quasars in the sample is 8 out of a total of 94 which gives a radio-loud fraction of $8.5 \pm 2.9$ per cent.   A binomial distribution was used to determine the error on this fraction.  To see if there was any evolution based on this fraction the value was compared to the work of \citet{jiang07}.  They looked at 30,000 optically selected quasars from the Sloan Digital Sky Survey matched to the FIRST radio survey.  From this data they have the radio-loud fraction as a function of both redshift and optical luminosity as given below:

\begin{equation}
\rm \log \left( \frac{RLF}{1-RLF} \right) = b_0 + b_z \log(1+z) + b_M (M_{2500}+26)
\end{equation}

where RLF is the radio-loud fraction, z is the redshift, $\rm M_{2500}$ is the absolute magnitude at 2500~\AA, b$_0 = -0.132 \pm 0.116$, b$\rm _z = -2.052 \pm 0.261$ and b$\rm _M = -0.183 \pm 0.025$.  The work of \citet{jiang07} only applies up to a redshift limit of \around 4.0 so our sample would be extrapolating their equation.

The left panel of Figure \ref{fig:plot_jiang07_mod}, using the equation of \citet{jiang07}, shows the radio-loud fraction as a function of redshift at $\rm M_{2500} = -28.38$, the average of the quasar sample.  The light blue is the error on this function.  The black point is the radio-loud fraction from our quasar sample at the average redshift.  The black lines cover the range of redshifts covered in our sample with the dotted lines showing the range of the error.  As can be seen the function from \citet{jiang07} and the value from our work lie within $1 \sigma$ of each other.  This indicates that our data is consistent with the equation of \citet{jiang07} in showing that for increasing redshift the radio-loud fraction decreases for ultraluminous quasars.   

The right panel of Figure \ref{fig:plot_jiang07_mod}, using the equation of \citet{jiang07}, shows the radio-loud fraction as a function of $\rm M_{2500}$ at redshift$= 4.67$, the average of the quasar sample.  The light blue is the error on this function.  The black point is the radio-loud fraction from our quasar sample at the average $\rm M_{2500}$.  The black lines cover the range of $\rm M_{2500}$ covered in our sample with the dotted lines showing the range of the error.  As can be seen the function from \citet{jiang07} and the value from our work lie within less than $1 \sigma$ of each other.  This indicates that our data is consistent with the equation of \citet{jiang07} in showing that for increasing optical luminosity the radio-loud fraction increases for ultraluminous quasars. 

One obvious criticism of this result is that we are using a rather large absolute magnitude range in our sample of $\rm M_{2500} = -30.00$ to $-$27.51 to do our comparison.  If we break the magnitude range into two bins ($\rm M_{2500} = -30.00$ to $-28.5$ and $\rm M_{2500} = -28.5$ to $-27.5$) we get 4 and 5 radio loud quasars out of 35 and 59 optical quasars respectively.  This gives a radio-loud fraction of $11.4 \pm 5.4$ per cent for the brighter range and $8.5 \pm 3.6$ per cent for the fainter range both of which are still in agreement with the equation of \citet{jiang07}.

{It is known that few galaxies at redshifts greater than \around 4 resemble present-day spiral or elliptical galaxies \citep{beckwith06}, and that  giant ellipticals or their progenitors are the main source of radio loud quasars \citep{floyd10,tadhunter16,rees16,rusinek20}.  If these galaxies represent a smaller fraction of galaxies at these redshifts, it suggests that the radio loud fraction of quasars should also decrease.   Thus our observed evolution of the radio loud fraction confirms current models of  galaxy evolution.}


\begin{table*}
\centering
\caption{{The radio loud fraction of high redshift quasars from the literature with their redshift ranges and the minimum absolute 2500 magnitudes of their sample.}}
\label{tab:literature}
\begin{tabular}{ccccc} 
\hline

\      & Radio Loud    & Redshift & Redshift    & Min Mag \\   
Source & Fraction (\%) & Min      & Max         & 2500 Abs \\

\hline
This work        & $8.5 \pm 2.9$ & 4.4 & 5.2 & -27.5 \\
\cite{yang16}    & $7.1 \pm 2.6$ & 4.7 & 5.4 & -27.0 \\
\cite{banados15} & $8.1^{+5.0}_{-3.2}$ & 5.5 & 6.4 & -26.9 \\
\cite{liu21}     & $7.1 \pm 2.7$ & 5.5 & 6.5 & -25.8 \\

\end{tabular}
\end{table*}



\begin{figure}
  \includegraphics[width=\columnwidth]{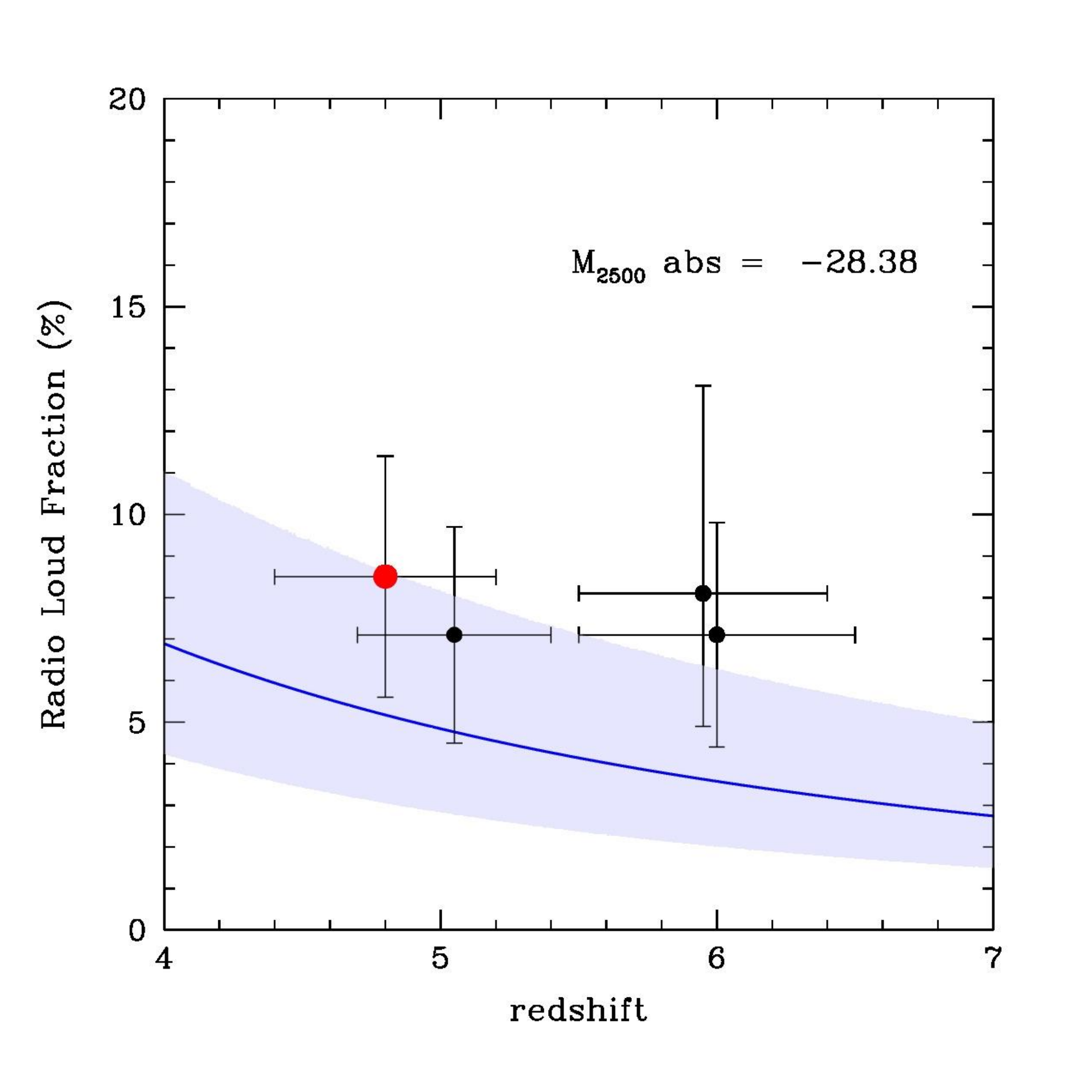}
  \caption{{The radio loud fraction of high redshift quasars from the literature.  The values for each can be found in Table~\ref{tab:literature}.  The value for this paper is the large red point.  The blue line with the error range around it is the value from the \citet{jiang07} function taken at M$_{2500}$ absolute magnitude -28.38, the average of this work's quasar sample.}}
  \label{fig:plot_literature}
\end{figure}

 
{Table~\ref{tab:literature} shows the radio loud fractions for quasars in the literature for high redshift samples.  These values have been plotted in Figure~\ref{fig:plot_literature} along with the \citet{jiang07} function taken at M$_{2500}$ absolute magnitude -28.38, the average of this work's quasar sample.  The values from \citet{liu21} come from their luminous sample, which  has a magnitude limit one magnitude fainter than the other samples.  It is also their all-radio sample value.  Taken as a whole these literature values indicate that at the magnitudes they probe there is little evidence for evolution in the radio loud fraction beyond redshift 5 though there is still reasonable agreement with the function of \citet{jiang07}.  However it should be noted that the \citet{jiang07} function presented in the figure moves from $4.8^{+3.2}_{-1.9}$ per cent at redshift~=~5 to $3.6^{+2.8}_{-1.5}$ per cent at redshift~=~6.  These values are consistent with no change, therefore at these redshifts there is little evidence for evolution even from the (extrapolated) function of \citet{jiang07}. Larger differences between the literature values and the function of \citet{jiang07} are seen at fainter M$_{2500}$ absolute value, which is particularly relevant for the \citet{liu21} sample.}

{It is not clear why the radio loud fraction appears flat near redshift~=~6.  This is unexpected as the number and type of {galaxies}
 that usually host radio loud quasars (ellipticals and their progenitors) are still decreasing at these redshifts.  This suggests a process that is affecting the radio properties of these distant quasars.  For example, since redshift~=~6 is approximately the end of the epoch of reionisation,  perhaps the gas supply for quasars is larger (or maybe just denser and more neutral) in this era so that quasars do not have to rely solely on mergers to support their radio emission.  More research needs to be done in this area both in larger and deeper quasar surveys at these redshifts and in theoretical models that predict the radio loudness fraction of quasars at these redshifts.}    


\section{Conclusion}
\label{Conclusion}

A sample of 94 ultraluminous, optical quasars from \citet{onken22} in the range $4.4<$redshift$<5.2$, with z~band magnitude $<18.7$, were matched against the radio continuum survey of RACS.  Ten quasars had radio detections in RACS of which eight are considered to be radio-loud.  The sample thus has a radio-loud fraction of $8.5 \pm 2.9$ per cent.  \citet{jiang07} modeled the radio-loud fraction as a function of redshift and absolute magnitude at rest frame 2500~\AA.  The radio-loud fraction we measure is consistent with an extrapolation of this function showing the predicted decrease with redshift and an increase with absolute magnitude.


\section*{Acknowledgements}

CAO was supported by the Australian Research Council (ARC) through Discovery Project DP190100252.

FDE acknowledges support by the Science and Technology Facilities Council (STFC), by the ERC through Advanced Grant 695671 ``QUENCH'', and by the UKRI Frontier Research grant RISEandFALL.

The national facility capability for SkyMapper has been funded through ARC LIEF grant LE130100104 from the Australian Research Council, awarded to the University of Sydney, the Australian National University, Swinburne University of Technology, the University of Queensland, the University of Western Australia, the University of Melbourne, Curtin University of Technology, Monash University and the Australian Astronomical Observatory. SkyMapper is owned and operated by The Australian National University's Research School of Astronomy and Astrophysics. The survey data were processed and provided by the SkyMapper Team at ANU. The SkyMapper node of the All-Sky Virtual Observatory (ASVO) is hosted at the National Computational Infrastructure (NCI). Development and support of the SkyMapper node of the ASVO has been funded in part by Astronomy Australia Limited (AAL) and the Australian Government through the Commonwealth's Education Investment Fund (EIF) and National Collaborative Research Infrastructure Strategy (NCRIS), particularly the National eResearch Collaboration Tools and Resources (NeCTAR) and the Australian National Data Service Projects (ANDS).

This paper uses data from the VISTA Hemisphere Survey ESO programme ID:179.A-2010 (PI. McMahon). Based on observations obtained as part of the VISTA Hemisphere Survey, ESO Program, 179.A-2010 (PI:McMahon). The VISTA Data Flow System pipeline processing and science archive are described in \citet{irwin04}, \citet{hambly08} and \citep{cross12}.

This publication has made use of data from the VIKING survey from VISTA at the ESO Paranal Observatory, programme ID 179.A-2004. Data processing has been contributed by the VISTA Data Flow System at CASU, Cambridge and WFAU, Edinburgh.

The ASKAP radio telescope is part of the Australia Telescope National Facility which is managed by Australia’s national science agency, CSIRO. Operation of ASKAP is funded by the Australian Government with support from the National Collaborative Research Infrastructure Strategy. ASKAP uses the resources of the Pawsey Supercomputing Research Centre. Establishment of ASKAP, the Murchison Radio-astronomy Observatory and the Pawsey Supercomputing Research Centre are initiatives of the Australian Government, with support from the Government of Western Australia and the Science and Industry Endowment Fund. We acknowledge the Wajarri Yamatji people as the traditional owners of the Observatory site. This paper includes archived data obtained through the CSIRO ASKAP Science Data Archive, CASDA (https://data.csiro.au).

The National Radio Astronomy Observatory is a facility of the National Science Foundation operated under cooperative agreement by Associated Universities, Inc. CIRADA is funded by a grant from the Canada Foundation for Innovation 2017 Innovation Fund (Project 35999), as well as by the Provinces of Ontario, British Columbia, Alberta, Manitoba and Quebec. 


\section*{Data Availability}

The optical quasar data used in this paper come from \cite{onken22} with additional quasar data from the Milliquas v7.1 \citet{flesch15} found at https://quasars.org/milliquas.htm

The RACS data used in this paper comes from CASDA which can be accessed through a TAP query of https://casda.csiro.au/casda\_vo\_tools/tap through the table AS110.racs\_dr1\_gaussians\_galacticcut\_v2021\_08\_v01.

The VLASS data used in this paper comes from the VLASS VizieR Epoch 1 Catalog which can be found at https://vizier.cds.unistra.fr/viz-bin/VizieR-3?-source=J/ApJS/255/30/comp

The NVSS data used in this paper came from the NVSS Source catalog browser which can be found at https://www.cv.nrao.edu/nvss/NVSSlist.shtml




\bibliographystyle{mnras}
\bibliography{quasars.bib} 


\bsp	
\label{lastpage}
\end{document}